\documentclass[11pt,twoside]{article}
\usepackage{asp2010,graphicx}

\resetcounters

\begin{document}

\title{Lesson learned from (some) recurrent novae}
\author{Elena Mason,$^1$ and Frederick M. Walters$^2$
\affil{$^1$STScI, Baltimore, MD 21218}
\affil{$^2$Stony Brook University, Stony Brook, NY 11794-3800}
}

\begin{abstract}
In this talk  we present early decline and  nebular spectra of the
recurrent  novae YY  Dor and  nova LMC  2009.  These  and a  few other
recurrent   novae   of  the   same   type,   share  similar   spectral
characteristics and evolution.  We will critically discuss those common
features  suggesting same  white  dwarf progenitor  and post  outburst
phases for all of them.
\end{abstract}

\section{Two ``new'' recurrent novae in the LMC}
Nova YY\,Dor (nova LMC\,2004) and nova LMC\,2009 are two recurrent novae (RNe) observed in the Large Magellanic Cloud. YY\,Dor recorded outbursts are those of 1937 (McKibben 1941) and 2004 (Liller et al. 2004, see also Bond et al. 2004). Nova LMC\,2009 was discovered in outburst by Liller (2009) who also suggested its recurrent nova nature due to the close match of its coordinate with those of nova LMC\,1971. 

\begin{figure}[!ht]
\centering
\includegraphics[width=9.5cm,angle=270]{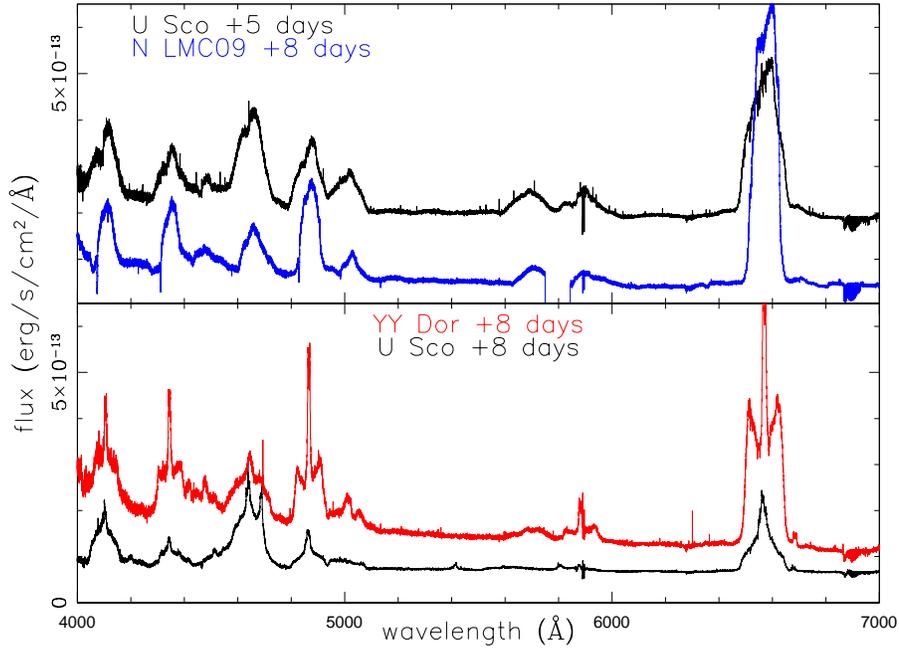}
\caption{The early decline spectra of YY\,Dor (bottom panel) and nova LMC\,2009 (top panel) compared with U\,Sco 2010 outburst spectra.}
\label{3max}%
\end{figure}

The two novae show very rapid declines (YY\,Dor $t_2$ and $t_3$ are $\sim$4 and $\gtrsim$10\,days, respectively; while for nova LMC\,2009 they are of the order of $\sim$5 and 11\,days, again, respectively) and broad emission lines in their maximum spectra (FWHM and FWZI of the order of $\sim$7000 and 10000\,km/s for YY\,Dor and $\sim$5000 and 6800\,km/s for nova LMC\,2009, respectively). Both these characteristics  and their repeated outbursts suggest that they belong to the class of U\,Sco type RNe (e.g. see Warner 1995, for an introduction about RNe classes). Indeed, their early decline spectra are remarkably similar to those of U\,Sco at similar phases, showing broad emission lines from He\,{\sc i}, N\.{\sc ii} and {\sc iii} in addition to the Balmer lines (Fig.\,\ref{3max}).

Here we present a few spectra of the two LMC novae and suggest that the U\,Sco type RNe represent a very homogeneous class of objects.  

\subsection{The nebular spectra}

\begin{figure}[!ht]
\centering
\includegraphics[width=9.5cm,angle=270]{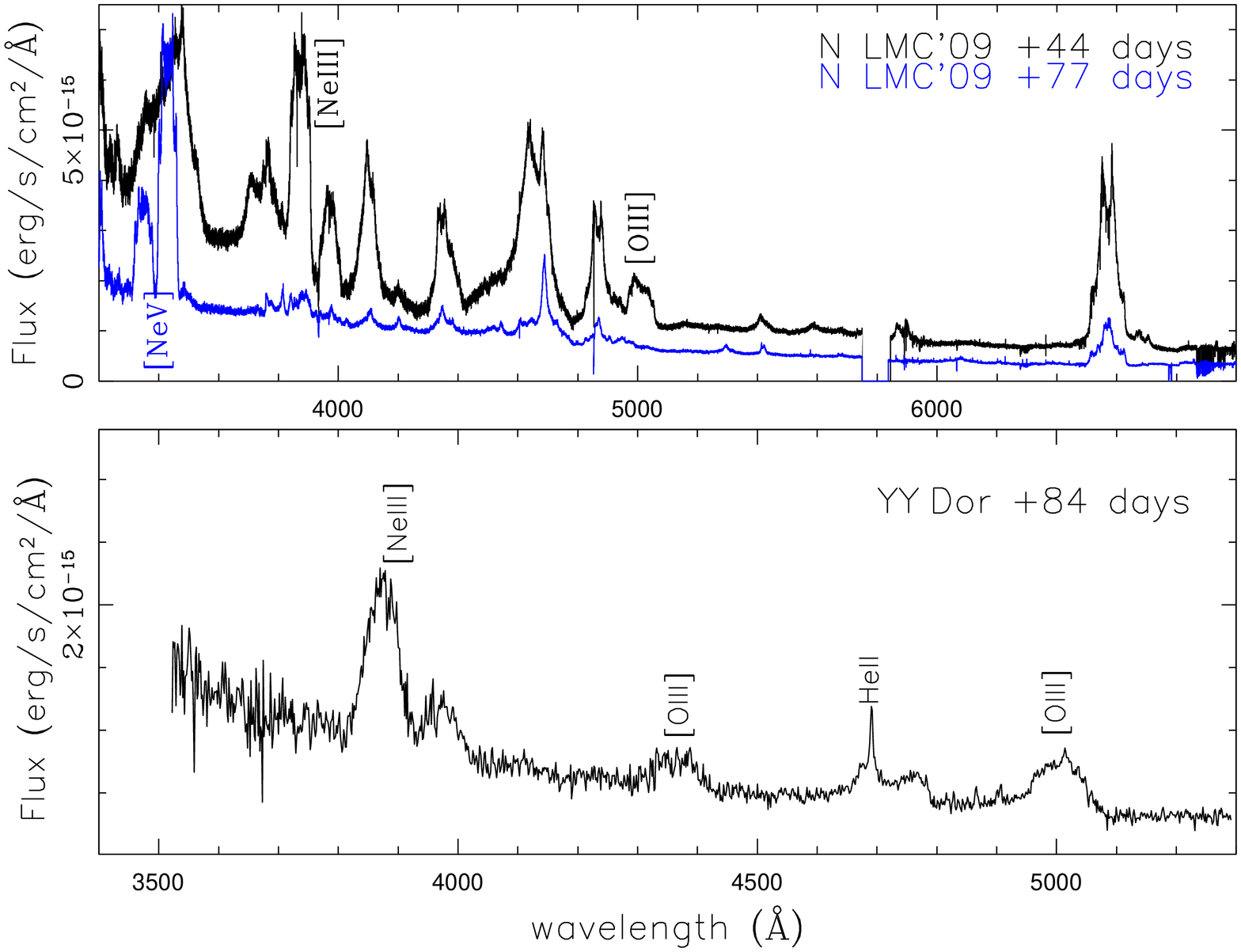}
\caption{The nebular spectra of YY\,Dor (bottom panel) and nova LMC\,2009 (top panel).}
\label{2neb}%
\end{figure}
Fig.\,2 shows the nebular spectra of YY\,Dor at day 84 (after maximum) and nova LMC 2009 at days 44 and 77. The two novae, similarly to U\,Sco, developed a nebular spectrum with transitions from [O\,{\sc iii}], [N\,{\sc ii}] as well as [Ne\,{\sc iii}]. In the case of nova LMC\,2009 also [Ne\,{v}] has been detected and it eventually became the strongest emission lines in the optical band (Fig.\,2). 

Mason (2011) analyzing the nebular spectra of U\,Sco computed the nova [Ne/O] relative abundance. She also showed that the [Ne/O] relative abundance in classical novae (CNe) can be used as a diagnostic for the composition of the underlying white dwarf (WD), i.e. to establish whether the CN progenitor is a CO WD or a NeO WD. In facts, CO WD CNe tend to have negative [Ne/O] relative abundances, while, ONe WD CNe have positive [Ne/O] ratios. 
Using the line flux of the emission lines [O\,{\sc iii}]$\lambda$5007 and [Ne\,{\sc iii}]$\lambda$3869 as in Mason (2011), we estimate the [Ne/O] relative abundance to be $\sim$1.98 and 2.07 for YY\,Dor and nova LMC\,2009, respectively. Unless the thermonuclear reaction in U\,Sco type RNe are somewhat different from those of typical CNe and in particular producing the so called CNO-breakout, the conclusions is that the two LMC RNe, as U\,Sco, host a ONe WD. We remark that current CN theory predict CNO breakout in cold slow accretors (T$_c<10^7$\,K, $\dot{M}<$10$^{-10}$\,M$_\odot$/yr, Glasner \& Truran 2009). These systems do not match the RN classes as they are expected to accrete at significantly larger rates ($\sim$10$^{-7}$\,M$_\odot$/yr, e.g. Hachisu \& Kato 2001). 

ONe WD, if accreting, are not expected to explode a supernova type-Ia (SN-Ia), once they reach the Chandrasekhar limit (e.g. Nomoto \& Kondo 1991). 
Hence, the three RNe, U\,Sco, YY\,Dor and nova LMC 2009, do not represent a viable SN-Ia progenitor. 

\subsection{The line profiles}

Fig.\,3 shows the evolution of the emission line profiles for the two LMC RNe. 
YY\,Dor line profiles are characterized by a double peaked narrow component of FWHM$\sim$700 km/s (decreasing with time), already in the maximum spectrum. Nova LMC\,2009 spectra developed a relatively narrow component (FWHM$\sim$1100-1200\,km/s), with time. 
As U\,Sco developed, too, a narrow component a few days after outburst, it is  interesting to compare these narrow emissions and check whether they could have a same origin. 

U\,Sco developed a narrow emission component in the H (Balmer, Paschen and Brackett series), He\,{\sc i} and He\,{\sc ii} lines (e.g. Mason et al. 2012) a few days after maximum. This narrow component was not observed in any of the nebular transitions at any time (see also Diaz et al. 2010); nor He\,{\sc ii}$\lambda$4686 broad emission was ever observed during U\,Sco decline. In addition, this narrow component appeared at about the same time the super soft source (SSS) phase started and displayed a radial velocity motion. Mason et al. (2012) interpreted U\,Sco narrow emission component and that of the He\,{\sc ii} line in particular, as a signature of the restoring accretion from the secondary star. 
This interpretation is further supported by Thoroughgood et al. (2001) time resolved spectroscopy obtained 49\,days after the 1999 outburst. Thoroughgood et al. (2001) clearly showed that the narrow He\,{\sc ii} emission forms an accretion disk and mirrors the WD orbital motion.   

We lack same quality time coverage and multi-band spectroscopic observations for the two LMC RNe, however, we can notice a few similarities. 
In particular, nova LMC\,2009 spectra display the narrow emission component only in the H  and He lines. The forbidden transitions and, in particular, the [Ne\,{\sc v}] show broad emission lines whose profile is identical to the broad component of the permitted transitions; however, they lack any narrow emission component. In addition, according to Schwarz et al. (2011) the RN enters its SSS phase between day 40 and 80 (though initially with low SNR), which is when we observe the appearance of the narrow component in our spectra. 

The case of YY\,Dor is even more uncertain and possibly different. First, there are no X-ray observations or monitoring in the literature. Second, we observed the narrow emission component already in  our first spectrum (day 5), in the H Balmer and O\,{\sc i} lines, $\lambda$7774 and $\lambda$8446 (but not in the He\,{\sc ii} which appeared only at day 16). The detection of the narrow emission in the O\,{\sc i} lines suggests that it originates in the ejecta, as only the ejecta structure and abundances can account for the detection of the low optical depth O\,{\sc i} emission lines against the larger optical depth continuum. Third, YY\,Dor narrow emission component does not display any significant radial velocity motion. Munari et al. (2011) and Shore et al. (2013) could reproduce narrow emission components superposed to broad emission line profiles by assuming bipolar ejecta and/or bipolar+equatorial disk ejecta.  

\begin{figure}
\centering
\includegraphics[width=5.5cm,angle=0]{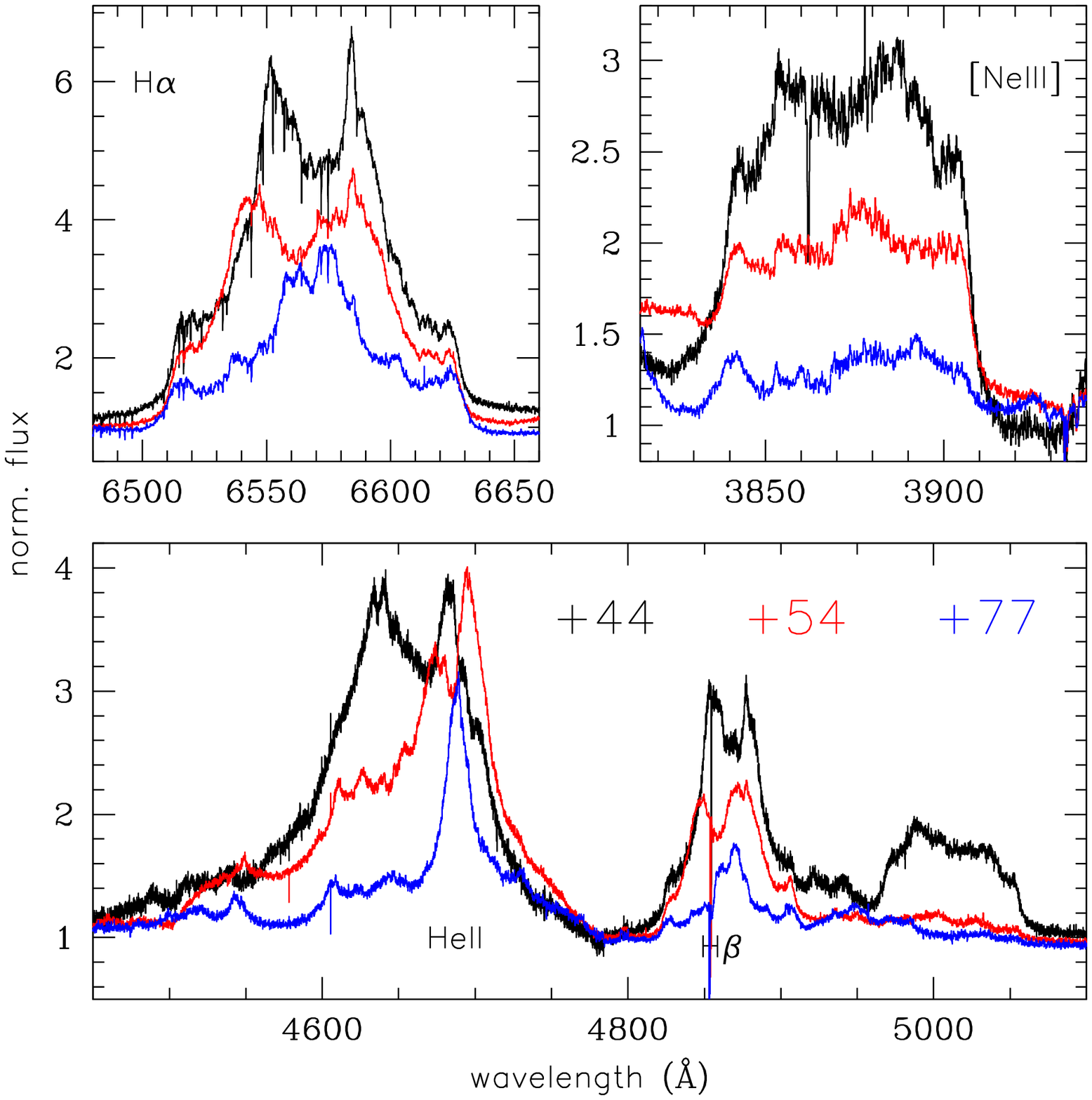}
\includegraphics[width=5.5cm,angle=0]{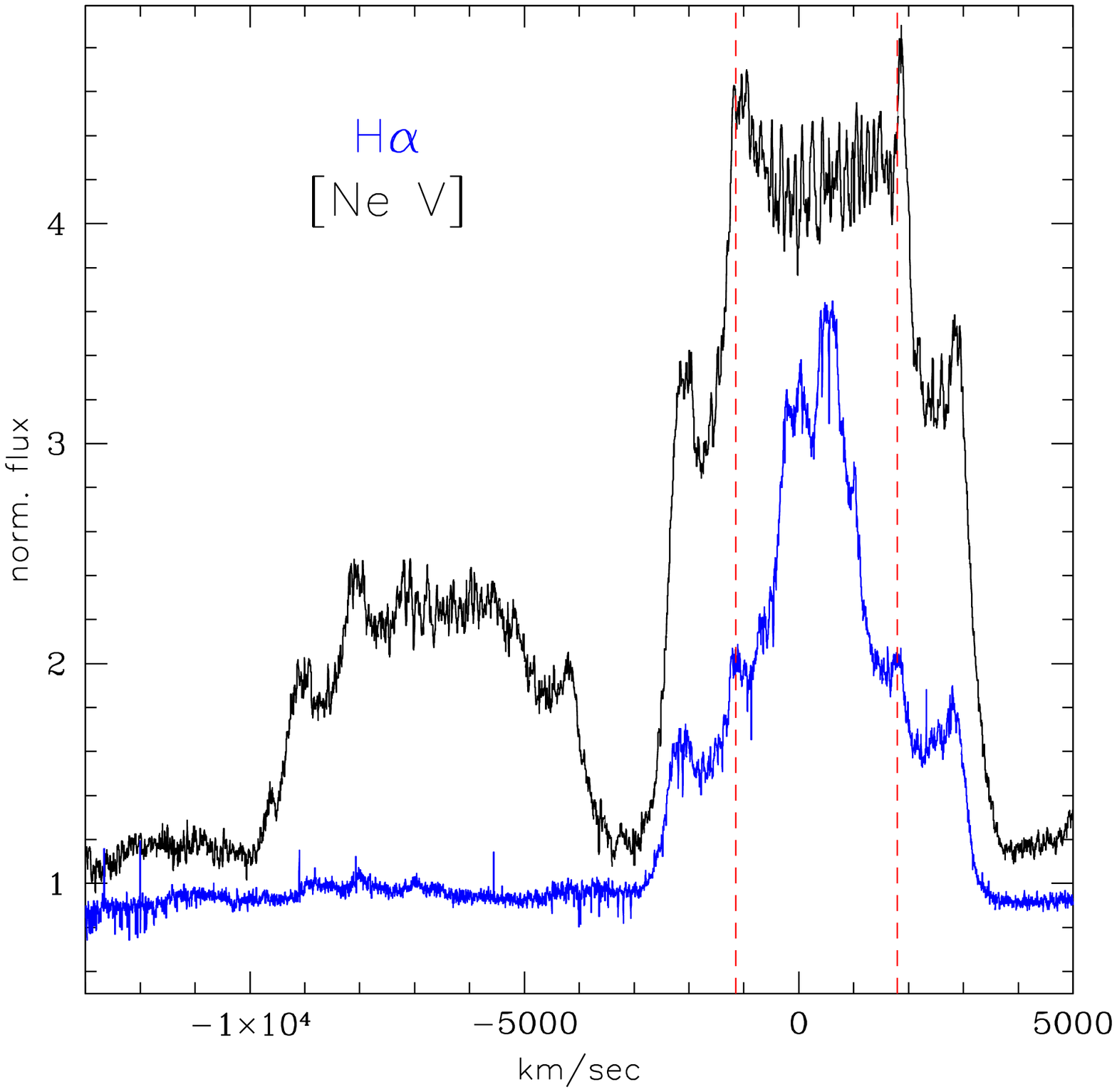}
\includegraphics[width=5.5cm,angle=0]{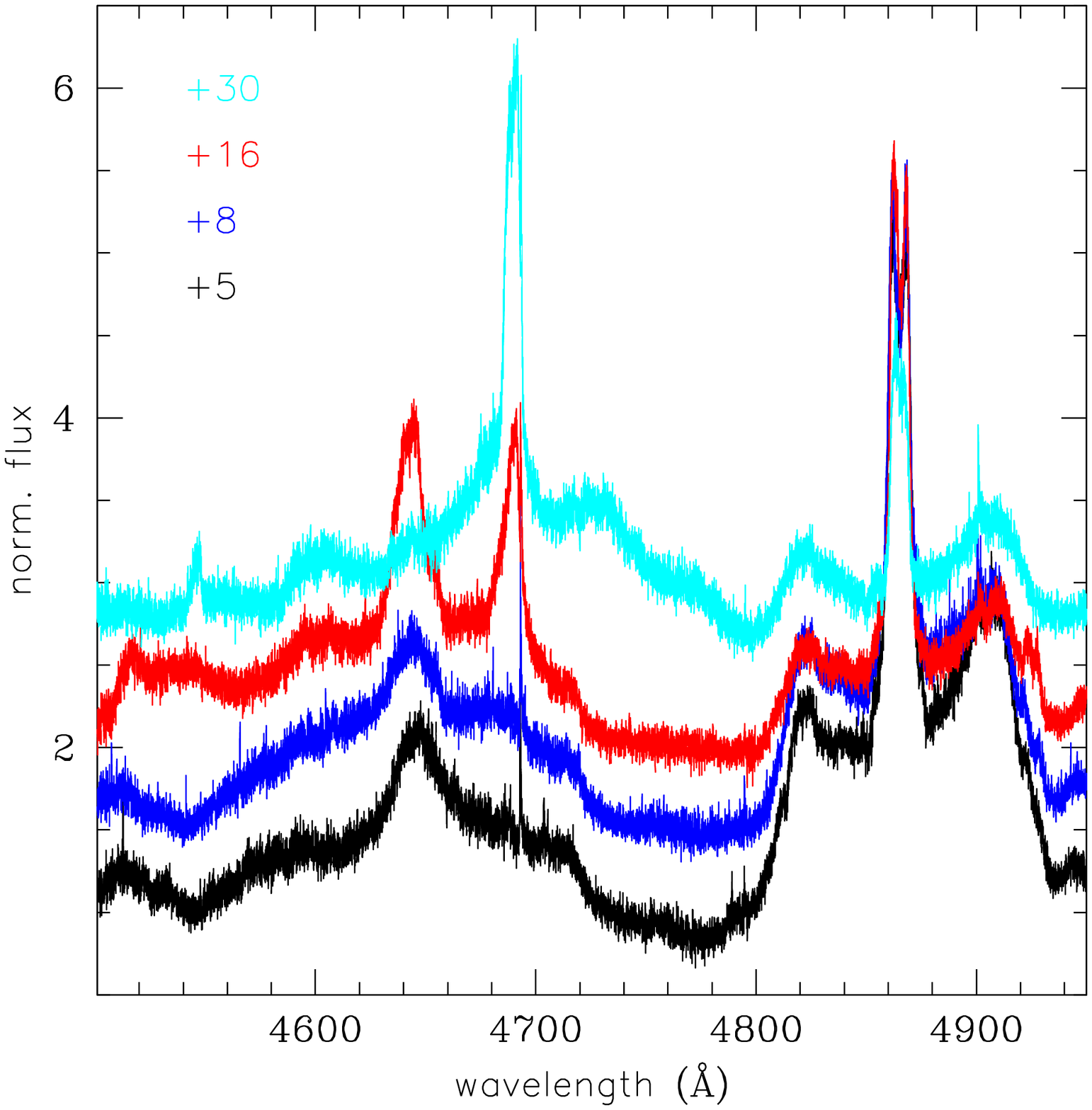}
\includegraphics[width=5.5cm,angle=0]{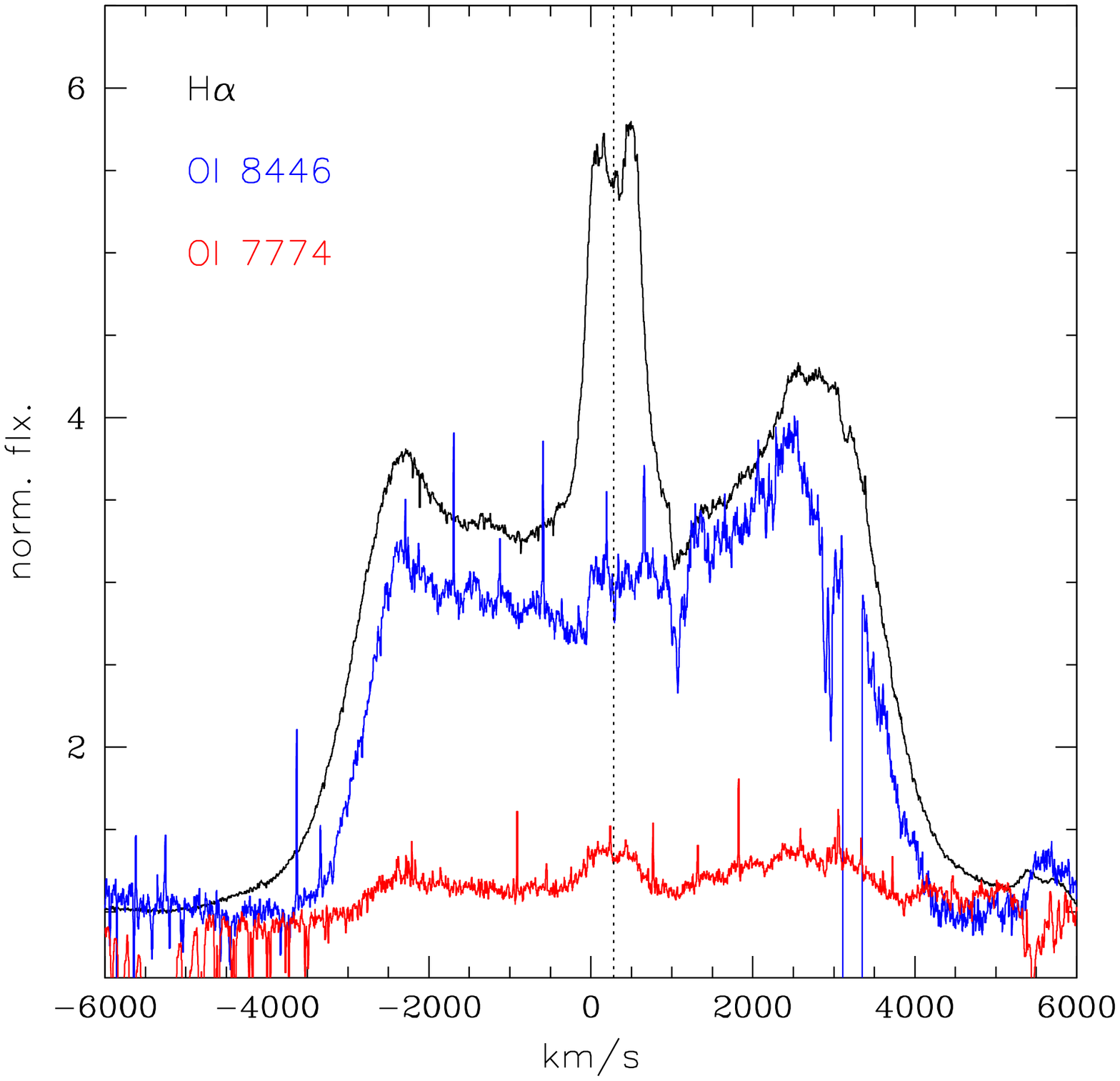}
\caption{The evolution of the line profiles in YY\,Dor (bottom left panel) and nova LMC\,2009 (top left panels). In YY\,Dor the He\,{\sc ii}$\lambda$4686 narrow emission appears ``only'' after 16 days from maximum. In YY\,Dor maximum spectrum (right bottom panel) the narrow component is detected also in the O\,{\sc i} line, suggesting an ejecta origin. In nova LMC\,2009 the narrow component (both He and H) starts to develop around day +54 and is well defined by day +77 (top left panels). Note how the nebular lines lack such a narrow component (top right panel).}
\label{2neb}%
\end{figure}

\section{The U\,Sco type recurrent novae}

In Table\,1 we list all the known RNe of the U\,Sco type (plus a couple of candidates such), together with their outburst year and a number of observational characteristics which allow us to directly compare them with U\,Sco, the prototype system. In particular, we report in column 3 and 4 the epoch (in days after maximum) of the first observed nebular spectrum and the last spectroscopic observation, respectively. For those objects which have an observed nebular spectrum we also report in column 5 their [Ne/O] relative abundance. The question mark in that same column simply means that the [Ne/O] relative abundance could not be calculated due to the lack of observed nebular spectra or of adequate spectral coverage. 
Hence, with the exception of the U\,Sco and the two LMC RNe discussed herein the other RNe of the U\,Sco class have not been followed into their nebular phase. The fact that three out of three U\,Sco type RNe for which nebular spectra exists show strong Ne emission lines with [Ne/O]$>$0, suggest that this characteristics is possibly common to all U\,Sco type RNe, i.e. all of them host a ONe WD primary. 
It will be important to follow up all their future outbursts well into the nebular phase to definitely prove the homogeneity of this class of RNe and rule it out (or not) from that of candidate SN-Ia progenitors. 

A quick inspection of the U\,Sco type RNe reveal that all of those with published spectra are characterized by the appearance of a narrow emission component with the He\,{\sc ii} emission line quickly overtaking, in strength, the H$\beta$ one. Therefore, we list in Table\,1 also the epoch of the appearance of the narrow emission component (H or any line, column 5), the appearance of the He\,{\sc ii} narrow emission component (column 6) and the start of the SSS phase (column 7). The colons indicates that the number and/or the detection are uncertain either because the spectrum is not available or because its resolution is too low. The dash, instead, indicates that the observation is missing. Despite the data are sparse and highly inhomogeneous, it seems that there is a communality of line profiles and, possibly, a connection between the appearance of the He\,{\sc ii} narrow emission at 4686\,\AA\, and the start of the SSS phase. While the SSS phase simply means that the burning envelope of the WD has become visible and the high energy photons can escape and cross the whole ejecta, the fact that the He\,{\sc ii} emission is always significantly narrow argues against its formation within the ejecta gas. In addition, should it form in the reforming accretion system within the primary Roche lobe, then the binary system is restoring accretion early during the outburst decline, with possible interesting implications on the ejection process (e.g. no or very short wind phase).
It will be important to adequately monitor future outbursts of the U\,Sco type RNe (as well as that of any CN displaying a narrow emission component) with coordinated optical and X-ray observations as well as time resolved optical spectroscopy thus to confirm the mentioned correlation and determine whether the narrow emission component shows any orbital/periodic radial velocity suggesting it arises from the restoring accretion and/or the reforming disk.  

\begin{table}
\caption{Observational properties of the U\,Sco type RNe, see text for details and explanations. Note that the RN CI\,Aql has not been included in the list as Ijiima (2012) has shown it does not belong to the U\,Sco type class. The RNe marked with the $\star$ are only candidate U\,Sco type RNe. }
\smallskip
\begin{center}
{\small
\begin{tabular}{cccccccccc}
\tableline
\noalign{\smallskip}
nova & outburst & 1$^{st}$ Neb. & last & [Ne/O] & n em. & n He\,{\sc ii} & SSS & REF\\
 & year & spc. & spc. &  &  & & & \\
\noalign{\smallskip}
\tableline
\noalign{\smallskip}
V394\,CrA & 1987 & - & 67 & ? & 20 & 20 & - & (a)\\
LMC 1990 N.2 & 1990 & - & 10-15 & ? & 8 & 8 & - & (b)\\
V2487\,Oph & 1998  & - & $>$3 & ? & $>$3: & : & - & (c)\\
YY\,Dor & 2004 & 41 & 92 & $>0$ & $<$5 & 16 & - & this work\\
V2672\,Oph $\star$  & 2009 & - & 8 & ? & 1-2 & - & 10 & (d) \\
LMC 2009 & 2009 & 44 & 77 & $>0$ & 54-77 & 54-77 & $\sim$50 & this work\\ 
U\,Sco & 2010 & 46 & 104,163 & $>0$ & 8 & 8 & 8-12 & (e) \\
KT\,Eri $\star$ & 2011 & 279 & 279 & ? & : & $\sim$65 & $\sim$60 & (f) \\
\noalign{\smallskip}
\tableline
\end{tabular}
}
\end{center}
\end{table}

\section{Summary \& conclusion}

From the analysis of the nebular spectra of the two LMC novae, YY\,Dor and nova LMC\,2009, we suggest that they both host an ONe WD similarly to U\,Sco. By observing their line profiles and those of same class RNe in the literature, we speculate that it could be possible that most of the systems are explained similarly to U\,Sco and that they recover accretion early during the outburst decline phase.   
In order to confirm these conclusions it is necessary to gather more well targeted observations and in particular 1) cover the nebular phase of U\,Sco type and all RNe, through UV and/or optical spectroscopic observations; 2) arrange coordinated observations in the X-band and optical spectroscopy (preferably mid to high resolution); 3) perform time resolved spectroscopy at designated decline phase.     

\acknowledgements EM is grateful to Robert E. Williams for the always interesting discussions and helpful feedbacks in the interpretation of the nova spectra. 

\vspace{0.2cm}
{\bf References}

Bond, H. E.; Walter, F.; Espinoza, J.; et al., 2004, IAUC 8424

Diaz M. P.; Williams R. E.; Luna G. J.; et al. AJ, 140, 1860

(c) Filippenko, A. V.; Leonard, D. C.; Modjaz, M.; et al. 1998, IAUC 6943

Glasner, S. A.; Truran, J. W., 2009, ApJ, 692L, 58

Hachisu, I.; Kato, M., 2001, ApJ, 558, 323

Iijima, T., 2012, A\&A, 544, 26

(f) Imamura, K.; Tanabe, K., 2012, PASJ, 64, 120

Liller, W.; Pearce, A.; Monard, L. A. G., 2004, IAUC 8422

Liller, W; 2009, IAUC 9019

(e) Mason, E., 2011, A\&A, 532, L11

(e) Mason, E.; Ederoclite, A.; Williams, R. E.; et al. , 2012, A\&A,  544, 149

McKibben, V., 1941, Harvard College Observatory Bulletin, 915, 1

(d) Munari, U.; Ribeiro, V. A. R. M.; Bode, M. F.; et al., 2011, MNRAS, 410, 525

Nomoto K., Kondo Y., 1991, ApJ, 367, L19

Schwarz, Greg J.; Ness, Jan-Uwe; Osborne, J. P.; et al., 2011, ApJS, 197, 31

(b) Sekiguchi, K.; Caldwell, J. A. R.; Stobie, R. S.; et al., 1990, MNRAS, 245, 28

Shore, S. N.; Schwarz, G. J.; De Gennaro Aquino, I.; et al, 2013, A\&A, 549, 140

Thoroughgood T.D.; Dhillon V.S.; Littlefair S.P.; et al., 2001, MNRAS, 327, 1323

Warner, B., 1995, in {\it Cataclysmic Variable star}, Cambridge University Press

(a) Williams, R. E.; Hamuy, M.; Phillips, M. M.; et al., 1991, ApJ, 376, 721

\end{document}